# NUANCE: NEAR ULTRASOUND ATTACK ON NETWORKED COMMUNICATION ENVIRONMENTS

Forrest McKee[1] and David Noever[2]
PeopleTec, 4901-D Corporate Drive, Huntsville, AL, USA, 35805
[1]forrest.mckee@peopletec.com     [2] david.noever@peopletec.com

## ABSTRACT

*This study investigates a primary inaudible attack vector on Amazon Alexa voice services using near ultrasound trojans and focuses on characterizing the attack surface and examining the practical implications of issuing inaudible voice commands. The research maps each attack vector to a tactic or technique from the MITRE ATT&CK matrix, covering enterprise, mobile, and Industrial Control System (ICS) frameworks. The experiment involved generating and surveying fifty near-ultrasonic audios to assess the attacks' effectiveness, with unprocessed commands having a 100% success rate and processed ones achieving a 58% overall success rate. This systematic approach stimulates previously unaddressed attack surfaces, ensuring comprehensive detection and attack design while pairing each ATT&CK Identifier with a tested defensive method, providing attack and defense tactics for prompt-response options. The main findings reveal that the attack method employs Single Upper Sideband Amplitude Modulation (SUSBAM) to generate near-ultrasonic audio from audible sources, transforming spoken commands into a frequency range beyond human-adult hearing. By eliminating the lower sideband, the design achieves a 6 kHz minimum from 16-22 kHz while remaining inaudible after transformation. The research investigates the one-to-many attack surface where a single device simultaneously triggers multiple actions or devices. Additionally, the study demonstrates the reversibility or demodulation of the inaudible signal, suggesting potential alerting methods and the possibility of embedding secret messages like audio steganography.*

## 1. INTRODUCTION

The present research examines high-frequency inaudible commands that affect network behavior. These novel attack surfaces open a gateway to significant financial, health, and personal information that may remain undetected for alerts and long monitoring periods. We explore this attack surface using inaudible commands (Figure 1). The work investigates the potential effects and vulnerabilities of near-ultrasonic attacks on voice-activated devices and seeks to identify possible countermeasures. The research focuses on mapping inaudible commands to consequential cases for cybersecurity when combined with digital signal-processing techniques.

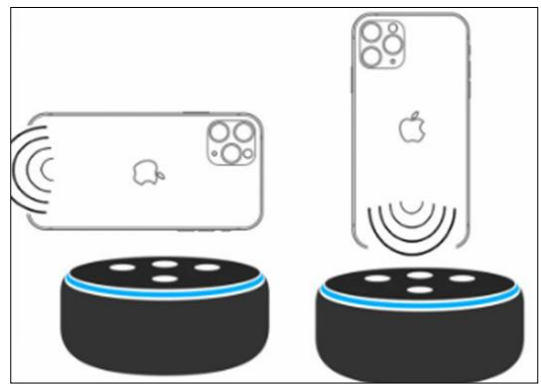

*Figure 1. Example inaudible signal from a smartphone broadcast to a home device receiver*

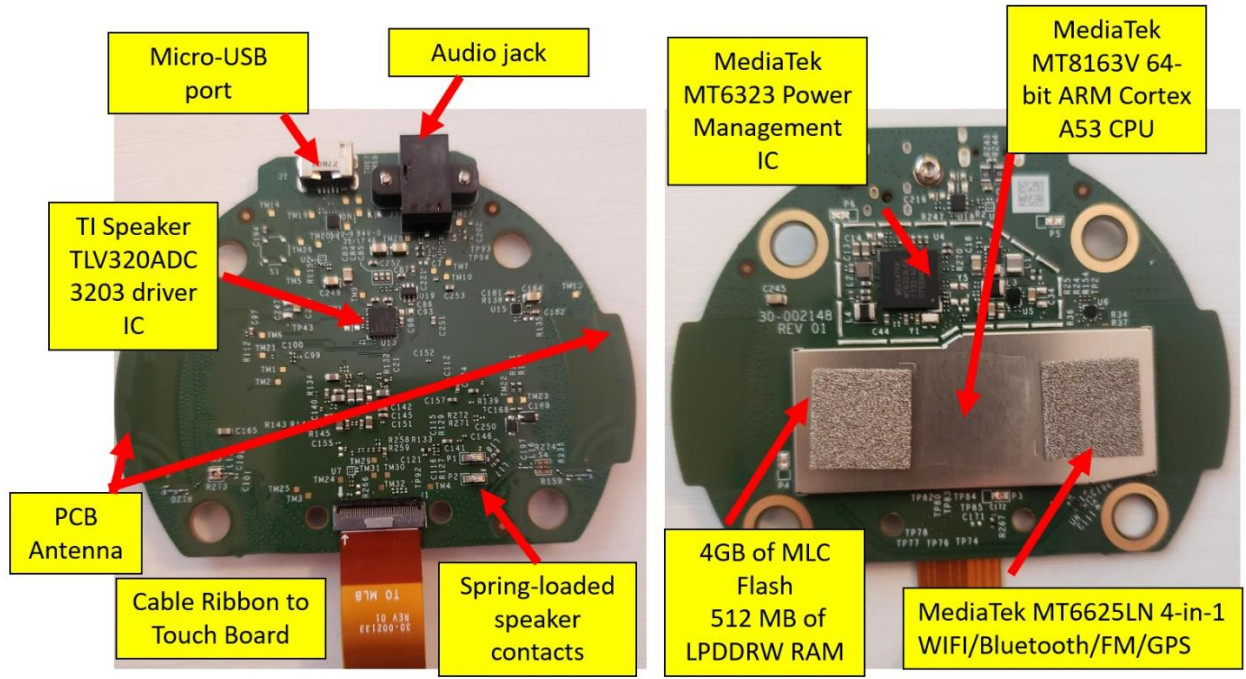

*Figure 2. Printed Circuit Board (PCB 1) from Echo Dot Generation 2 Teardown Highlights the Analog-Digital Interfaces*

Voice-activated devices permeate workplaces, homes, and smartphones, such that at any time of day, a listening microphone likely sits within arm's reach [1-2]. The growing prevalence of voice-activated devices [1-2] has increased security risks, with ultrasonic attacks posing a significant novel threat. Industry surveys reveal that in 2023, 4.2 billion digital voice assistants will exist worldwide (or about 1 for every two people globally), with an annual expected growth of 100% annually to 8.4 billion in 2024 [2]. As the most popular smartphone providers, Apple Siri and Google Assistant account for 36% of users each (approximately 1.5 billion) [1]. In user engagement terms, 3.1 billion voice searches occur each month [1], including in higher-stress environments since 62% of US smartphone users employ voice assistants while driving. Amazon's Alexa was the top-rated voice command platform in 2020 [1], dominating the US market with a 70% market share.

Despite this saturation, two of three surveyed smart speaker "non-owners" do not purchase because the device is always listening [1], while a similar percentage of owners think voice-operated devices with always-on capability conveniently make their life easier. As part of understanding the scope, 56% of people who use a voice assistant keep the device using open microphones or turned on all the time [1]. Audio security is the primary concern for 42% of voice shoppers in the US (or 44 million devices extrapolated to 1.8 billion devices globally) [1]. Researchers [3-17] have made progress in developing defense mechanisms and detection techniques to counter these attacks, but continued vigilance and innovation are required [14] to ensure the safety and privacy of users. By staying informed of new attack vectors [15] and refining existing defenses [11], the cybersecurity community can help mitigate the risks associated with voice-activated devices [10, 18] and create a more secure environment for their use [16].

Consumer acceptance of an open microphone in private and public spaces suggests concerns about securing these devices. Typically, suppose the microphone activates using a wake word [20] ("Hey Google," "Alexa" [10], "Siri" [21], "Cortana"). In that case, there is no other security [20,22-24], countermeasures [6], or identifier [25-27] required to access what might offer much deeper access to one's personal finances, health records, or credit cards. Researchers have demonstrated the potential for abusing and exploiting the surrounding microphones to accomplish complex tasks [19, 28-32], often without user input or knowledge [19,33-49].

The pioneering work [19], called Near Ultrasonic Inaudible Trojan (or NUIT), features a matrix of device combinations that succumb to microphone nonlinearities. Perhaps most impressively, the authors show novel attacks that require no special hardware (unlike Dolphin Attacks [9]), although the spatial proximity between the device and target appears critical to NUIT's success. They generalize two cases for single devices (NUIT1 self-attack) and dual devices (NUIT2, mixed devices which broadcast and receive near the source and target). We anticipate in this research a host of one-to-many attacks (NUIT-N) which further generalize the attack surface and share some standard features of viruses or worms that penetrate larger device populations simultaneously. Unlike the broad survey of many devices, the present approach addresses a single attack vector between phones and home devices (phone-to-Alexa) to evaluate the effects of specific commands and highlight the complex stacking of multiple interactive or persistent control of devices.

Unauthorized access to these devices has real-world consequences [50-59], whether manipulating a home automation system [60] or autonomous vehicles [61-63], compromising personally-identifying-information (PII), or uncredentialed access to bank accounts using default settings. One metric for assessing the size of the audio attack surface [10,11,19] might be the number of devices, their types, and typical locations in homes, environmental controls, and phones [19,64-70]. In addition to offensive cyber risks [71], the use of audio steganography of hidden or subliminal messages [72], covert channels [69,73-74], air-gap computer trojans [75], jammers [76], and firewalls [77] have received recent attention. Table 1 summarizes cross-device examples of open microphones that listen perpetually for the next instruction and second-order skills that enable complex tasks to commence from the wake word and command sequence.

*Table 1. Cross-device networked attacks by study references*

| Source/Target | iPhone | Alexa | Cortana | Android | YouTube | Home |
|---|---|---|---|---|---|---|
| **iPhone** | [19,21,32] | [10,19,44] | [19] | [19] | [19] | [19,28] |
| **Alexa** | [19,21] | [10,19,44] | [10,19,44] | [10,19,44] | | [19,28] |
| **Cortana** | [19,21] | [10,19,44] | [19,31] | | | [19,28] |
| **Android** | [19,21] | [10,19,44] | | [19] | | [19,28] |
| **YouTube** | [19,21] | [10,19, 44] | | | [19] | [19,28] |
| **Home** | | | | | | [19,28] |

Previous research has proposed novel methods and tests for inaudible [57] or covert commanding of voice-activated devices [4-16]. These approaches include attacks at ultrasonic frequencies beyond adult human hearing [8,19,21,29,32,38-40,43,45,47-48,52,60], embedded obfuscations [26,64], physical access and vibration or power lines [33,51,54,66], and synthetic voice or imitation methods [9,38,41.47]. In recent years, the rapid growth of voice-activated devices, such as Amazon's Alexa [10,19,21,44], has raised concerns about the security and privacy of these systems [46]. Ultrasonic attacks, which use inaudible sound frequencies to manipulate voice assistants, have emerged as a significant threat to the security of these devices [19].

Several studies have investigated the feasibility and effectiveness of ultrasonic attacks on voice-activated devices. The Dolphin Attack [9,38,41,47], for example, demonstrated that inaudible voice commands could be used to control popular voice assistants with external speakers capable of broadcasting and amplifying high-frequency signals. At the same time, other research explored the long-range capabilities of such attacks [19]. A recent study on Surfing Attacks [66] showed that ultrasonic-guided waves could interact with voice assistants covertly. These findings highlight the need for robust defense mechanisms against inaudible voice command attacks [4-17].

Various defense techniques have been proposed to counteract the threat of ultrasonic attacks. One approach involves using the physical fingerprinting of ultrasonic sensors to enhance their security [68], while another focuses on canceling inaudible voice commands against voice control systems [8]. For instance, an Ear Array [41] defends against Dolphin Attacks through acoustic attenuation. These defense mechanisms protect voice-activated devices from malicious ultrasonic interference [55]. In addition to defense mechanisms, researchers have developed techniques to detect ultrasonic attacks. Watchdog, a system proposed in one study [60], is designed to detect ultrasonic-based inaudible voice attacks on smart-home systems. Another detection method, FOCUS [74], is frequency-based and aims to identify covert ultrasonic signals in ICT (Information Control Systems). These detection techniques attempt to counter unauthorized access to voice-activated devices.

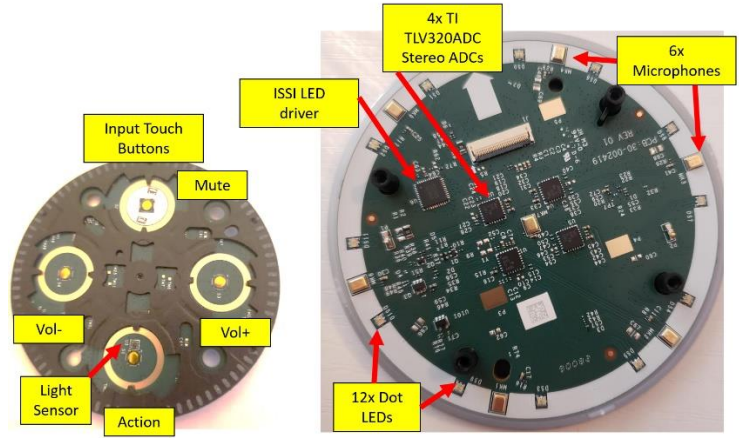

*Figure 3. Printed Circuit Board (PCB 2) on Echo Dot 2 Highlights the Sensory Inputs for Voice and Touch Commands*

Despite the advancements in defense and detection techniques, new attack vectors continue to emerge [10,19,44]. For example, the Near-Ultrasonic Covert Channels method utilizes software-defined radio techniques to create covert channels for transmitting data [69]. Similarly, the HVAC attack method evades classifier-based defenses in hidden voice attacks [13]. Ultrasonic attacks are not limited to voice-activated devices; they also target other systems, such as vehicular sensors [61-62, 64] and autonomous vehicles [62]. These researchers highlight the broader impact of ultrasonic attacks on various connected systems, particularly when the attack couples to a device capable of performing many dangerous downstream commands or executing voice-activated skills [28].

As shown in Figure 4, the present work explores a primary inaudible attack vector using Amazon Alexa voice services as the target and multiple input devices to deliver near-ultrasound trojans [19]. In contrast to previous approaches, the research explores the attack surface of one target device rather than surveying the large search space of voice-activated device combinations. The interest centers on characterizing the attack surface and exploring the practical implications of issuing inaudible voice commands to the Amazon Alexa system in various contexts and connected states.

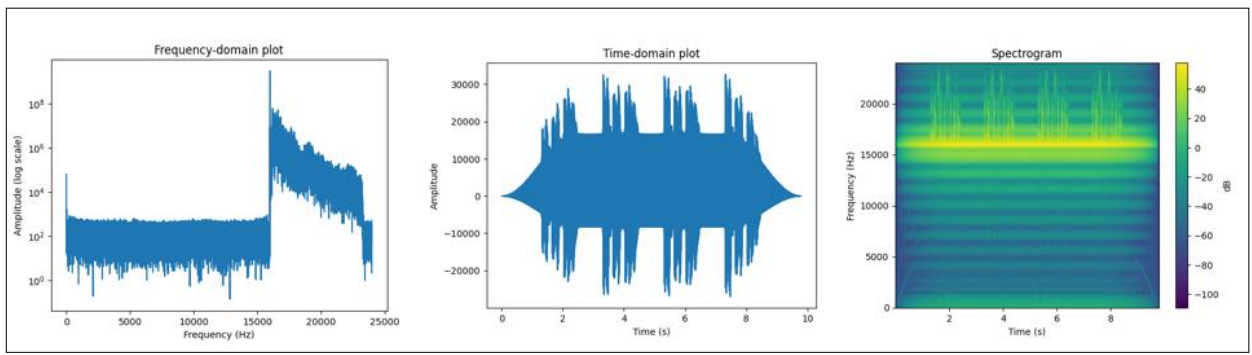

*Figure 4. SUSBAM signal in time and frequency domains*

To frame the problem systematically, we map each attack vector to an existing tactic or technique derived from the MITRE ATT&CK matrix [78-81]. Since the voice-activation element may appear on enterprise (Cortana), mobile (Android, iPhone), home, and Internet of Things (IoT) devices (Alexa, Google), many of the attack vectors appear across different frameworks. Thus, the MITRE ATT&CK framework included enterprise, mobile, and Industrial Control System (ICS) notation [82]. One advantage of this systematic approach is stimulating previously unaddressed attack surfaces that transfer from the accumulated (often trial-and-error) tactics and techniques learned from the cybersecurity world outside of voice activation. This approach ensures substantial coverage for detecting and designing new attacks. A second advantage involves pairing each ATT&CK Identifier with at least one plausible and tested defensive method since MITRE frameworks give both red (attack) team tactics and blue (defender) team tactics as prompt-response options [82].

## 2. METHODS

As first reported [19], the attack generates near-ultrasonic audio from audible sources using Single Upper Sideband Amplitude Modulation (SUSBAM). As illustrated in time and frequency space (Figure 5), the method transforms a spoken command to a voice-activated device using the modulated audio signal converted into a frequency range (16-22 kHz) beyond human-adult hearing. Stepwise, the method applies a low-pass filter with a cutoff of 6 kHz to the original spoken signal and removes components not essential to the SUSBAM generation. The generator normalizes the remaining audio signal and applies a carrier frequency (16 kHz) to shift the audible range of the input signal to a higher frequency (near ultrasonic, 16-22 kHz).  Figure 4 shows the original and transformed signal as an example in waveform (time domain) and spectral (frequency and power density domain). Figure 5 overlays the audible and inaudible signals as spectrograms.

To modulate the voice command, a 90-degree phase-shifted version gets created from a Hilbert-transformed input, then modulated as a combination of the original signal, its Hilbert-transformed version, and the assigned carrier frequency. To taper the signal's edges to zero smoothly, a Tukey window is applied to the modulated signal, which reduces spectral leakage in the frequency domain and prevents artifacts in the near-ultrasonic output. The method normalizes the SUSBAM signal to a 16-bit integer PCM audio format range, then casts the normalized SUSBAM signal to the int16 data type expected from an uncompressed WAV file. Compression in later stages of SUSBAM generation to MP3 formats loses functionality. In summary, the NUIT process reads an input audio signal, applies a low-pass filter, modulates the signal using SUSBAM to shift it to a near-ultrasonic frequency range, applies a Tukey window, normalizes, converts the signal to an integer format, and writes the processed signal to an output WAV file.

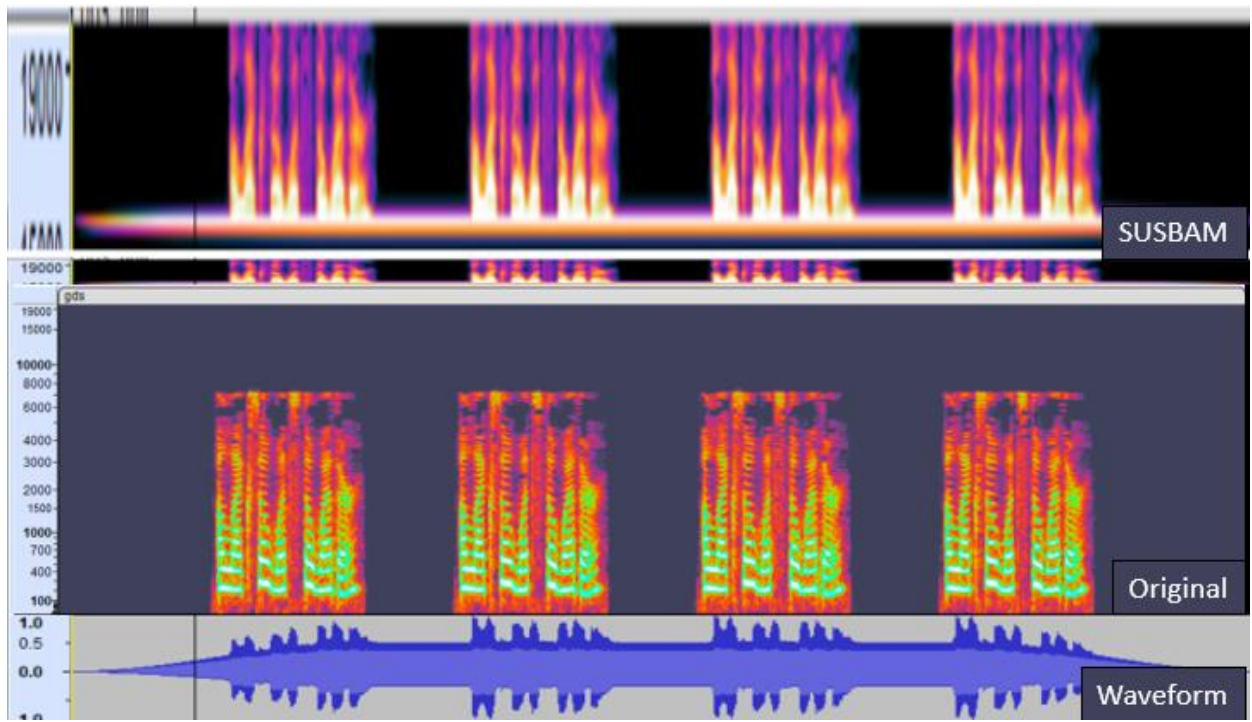

*Figure 5. Original (middle) and transformed (lower and upper) signals in time and frequency domains*

## 2.1. Exploring the Command Attack Surface

The MITRE framework represents a catalog for examining the available attack and defending surfaces for the near-ultrasound vectors. The Appendix outlines example stages grouped by tactics such as initial access, then the various techniques used in the literature. The most apparent result of an inaudible attack on voice-activated devices follows initial access using a drive-by compromise (Table 2). The attacker does not need keyboard availability (either locally or remotely) at the time of the initial entry. The attacker can access a set command list either innately or through additional applications (called "skills" in Alexa). A remote attacker may embed the vector in the silence of an online video (YouTube) that is currently popular or "going viral," then gain both a distribution "worm-like" quality along with accessing what is, in practice, a password-less means to discover financial, health, personal or other damaging information about the victim. The recommended MITRE defense for the attack is user training, which in this case might specialize to the position of the Echo device in a space accessible to the attacker or include specific profile configurations that limit what an unrecognized voice can accomplish.

Similarly, an attacker may use an ultrasonic attack to target valid accounts by employing ultrasonic signals to inject a malicious skill onto a target device, replacing an existing skill [82]. This tactic allows the attacker to prompt the user to enter or re-enter personal information, facilitating input capture. Upon successfully capturing the desired information, the attacker can then use it to access a valid account, which may have elevated privileges, enabling the attacker to move laterally within the target network. The recommended MITRE defense for the attack would be User Training and Access control. In practice, this could be to have a user verify a skill is genuine and to allow users to have the minimum privilege needed.

*Table 2. Example Near Ultrasonic Attack and the Paired Defend Tactics and Techniques*

| ATT&CK Tactic | ATT&CK Technique | D3FEND Tactic | D3FEND Technique | Ultrasonic Attack |
|---|---|---|---|---|
| **Initial Access** | T1189: Drive-by Compromise | User Training | D3-T1023: Security Awareness Training | Yes |
| **Privilege Escalation** | T1078: Valid Accounts | Access Control | D3-T1021: User Account Management | Yes |
| **Credential Access** | T1056: Input Capture | User Training | D3-T1023: Security Awareness Training | Yes |

Appendix A shows the plausible mapping of these attacks to known stages of the MITRE framework, along with paired defenses. A notable aspect of this map suggests new avenues that have not appeared in previous broad surveys meant to explore the various device and microphone combinations. Given the right skills, connectivity, and access, this work centers on the deeper attack surfaces available on one device.

## 2.2. Examining Text-to-Speech Engines

Text-to-Speech (TTS) has the advantage of being quickly generated at a large scale. TTS offers an experimental platform for testing NUIT commands as reproducible inputs, thus limiting any errors due to fluctuations in command delivery from the human voice. The research explored multiple candidates for TTS in trial-and-error experiments but ultimately selected gTTS. We did not initially find a significant effect for different voice synthesizers based on the age or gender of the TTS avatar. Future opportunities for TTS engines offer higher sample rates and more advanced generation techniques to aid Voice Control System (VCS) detection. Different gendered voices, accents, speed, pitch, and volume are all potential avenues to explore to increase NUIT reliability.

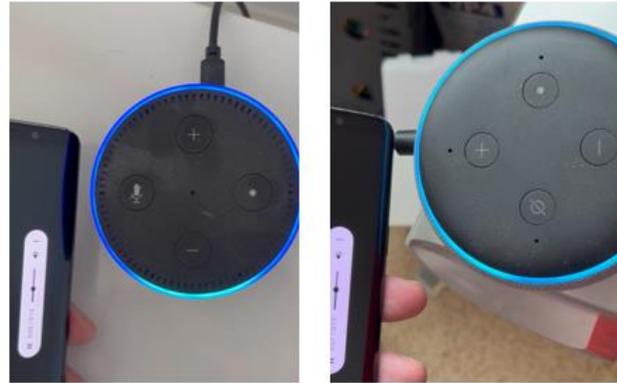

*Figure 6. NUIT2 Attacks from Android to Echo Dot Gen 2/3 Devices: "Alexa, what's the weather?"*

## 3. RESULTS

### 3.1 Survey Test of Voice-Activated Commands

Table 3 shows the success rate with unprocessed TTS commands and transformed TTS commands. The experiment included fifty orders to survey the attacks' effectiveness based on a repeatable text-to-speech generated input modified by the modulation algorithm. The unprocessed commands had a 100% success rate, while the processed ones had a success rate of 58%. 26% of the audio recordings got Alexa to acknowledge the command, but the device's microphone failed to recognize the wake word or could not follow the order. Alexa failed to trigger the final 16%. Figure 6 shows a two-panel image of an Android (S9 model) playing the "What's the weather?" attack on both Amazon Alexa Echo Dot Generation 2/3 devices. These attack vectors would qualify as NUIT2 [19], signifying that there is a two-device audible connection, one from the broadcaster (phone) to a receiver (home device).

More research is needed to diagnose why some commands had higher success than others, but a simple explanation would be that the VCS more easily understands some combinations of syllables from the TTS service than others. When these command-specific variables combine with the exploited nonlinearity in the VCS microphone may account for the lower success rates of the VCS algorithms.

*Table 3. Survey of Actionable Commands in the Voice-Activated Device and Success Rates for Recognition and Response as Actions*

| ID | Command | Original (mp3) | | | Near Ultrasound (NUIT) | | |
|---|---|---|---|---|---|---|---|
| | | fail | trigger | success | fail | trigger | success |
| 0 | Help | | | x | | | x |
| 1 | Mute | | | x | x | | x |
| 2 | Unmute | | | x | | | x |
| 3 | Stop | | | x | | x | x |
| 4 | Louder | | | x | | | x |
| 5 | Set the volume to five | | | x | | x | x |
| 6 | Play some music | | | x | | | x |
| 7 | Set a timer for one minute | | | x | | | x |
| 8 | What's playing? | | | x | | | x |

| ID | Command | Original (mp3) | | Near Ultrasound (NUIT) | |
|---|---|---|---|---|---|
| 9 | When is Christmas next year? | | x | | x |
| 10 | What's on my calendar for tomorrow? | | x | | x |
| 11 | What's in the news? | | x | | x |
| 12 | What's the weather like? | | x | | x |
| 13 | What's the traffic like? | | x | | x |
| 14 | What movies are playing? | | x | | x |
| 15 | What is Tom Holland's latest movie? | | x | x | |
| 16 | Who is in The Rolling Stones? | | x | | x |
| 17 | What's five plus seven? | | x | | x |
| 18 | Flip a coin | | x | | x |
| 19 | Pick a number between one and ten | | x | | x |
| 20 | What's the definition of ultrasound? | | x | | x |
| 21 | How do you spell Apple? | | x | | x |
| 22 | Did the Lakers win? | | x | x wrong cmd | |
| 23 | When do the Lakers play next? | | x | | x |
| 24 | Which profile is this? | | x | x | |
| 25 | What kid's skills do you have? | | x | | x |
| 26 | Wikipedia ultrasound | | x | | x |
| 27 | How tall is Steph Curry? | | x | x | |
| 28 | Tell me a joke | | x | | x |
| 29 | Beam me up | | x | | x |
| 30 | Set phasers to kill | | x | x | |
| 31 | Tea. Earl grey. Hot. | | x | x | |
| 32 | My name is Inigo Montoya | | x | | x |
| 33 | I want the truth | | x | x | |
| 34 | Party on, Wayne. | | x | x wrong cmd | |
| 35 | Show me the money! | | x | | x |
| 36 | What's the first rule of Fight Club? | | x | | x |
| 37 | Surely you can't be serious | | x | x | |
| 38 | Are you Skynet? | | x | x | |
| 39 | Party time! | | x | x wrong cmd | |
| 40 | Open the pod bay doors. | | x | | x |
| 41 | What is your quest? | | x | x wrong cmd | |
| 42 | Don't mention the war | | x | x | |

| ID | Command | Original (mp3) | | | Near Ultrasound (NUIT) | | |
|---|---|---|---|---|---|---|---|
| 43 | What is your cunning plan? | | | x | | x wrong cmd | |
| 44 | What is the loneliest number? | | | x | | x | |
| 45 | What is the best tablet? | | | x | x | | |
| 46 | Do aliens exist? | | | x | | x | |
| 47 | Where do you live? | | | x | x | | |
| 48 | How tall are you? | | | x | | | x |
| 49 | I think you're funny | | | x | | | x |
| | | | | | | | |
| | Total | 0 | 0 | 50 | 8 | 13 | 29 |
| | % | 0 | 0 | 1 | 0.16 | 0.26 | 0.58 |

A critical element of the NUIT design follows from eliminating the lower sideband to achieving a 6 kHz minimum from 16-22 kHz while remaining inaudible after transformation. In other words, the selective removal of modulated elements that might leak provides the spectral room to span the minimum detection needs of the non-linear microphone interactions. Figure 6 shows a two-device attack surface (NUIT2) with the broadcaster (Android) near the receiver (Alexa Echo) in two different models. The inaudible WAV file plays four times ("Alexa, what's the weather?"), and the blue ring on the target device signifies the two stages of acknowledgment (signal recognition) and the intended action (comprehension and result).

### 3.2 Extensible Attack Surfaces: NUIT-N Design

To extend the previous single and dual-device attacks [19], we outline a one-to-many design common to many public events where an audible broadcast might embed an inaudible command within the silence and effectively trigger an entire crowd. Examples include a public announcer (PA system), a live online broadcast (YouTube or Facebook live), a concert hall, or a network of home devices (Alexa groups). This version extends the one-device and two-device attack surface to any number we subsequently call NUIT-N (Near Ultrasonic Inaudible Trojan-N). We demonstrate this phenomenon in a straightforward setup as one broadcasting device (phone) that inaudibly controls multiple home devices simultaneously. In Figure 7, the NUIT signal sends "Alexa, what's the weather?" to both devices inaudibly, thus fulfilling the (N) requirement at a basic level. Video demonstration replays are available online [84].

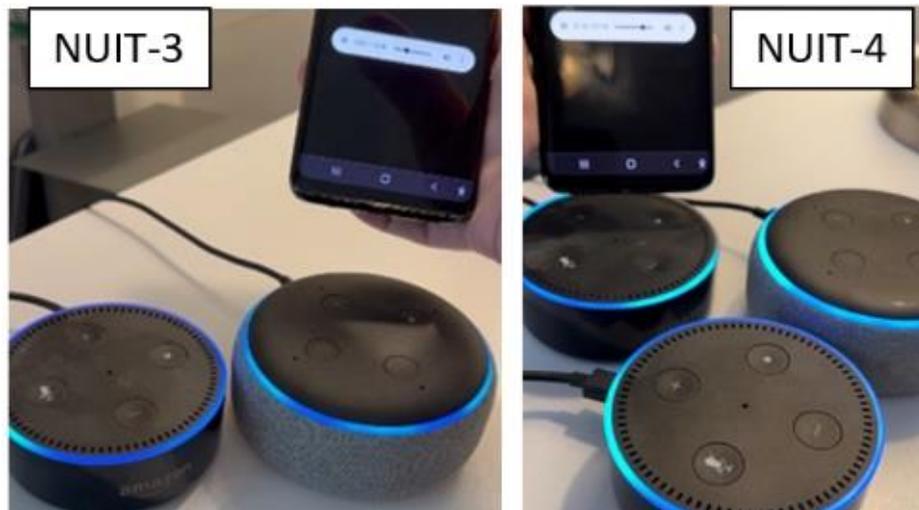

*Figure 7. NUIT-N Attack Linking a Single Broadcasting Device to Multiple Receivers Simultaneously*

### 3.3 Demodulation

Running the demodulation algorithm on the attack signal captured by an inexpensive headphone microphone reveals reversibility in ambient noise (Figure 8A). The slight degradation in the signal quality appears in both the spectrogram and waveform representations of the signal. The spectrogram shows that the demodulation recovers energy around 3kHz, while the digital signal recovers roughly 6kHz (Figure 8B). Aside from the ambient noise, a potential factor for the loss in signal quality is the speaker and microphone's frequency response.

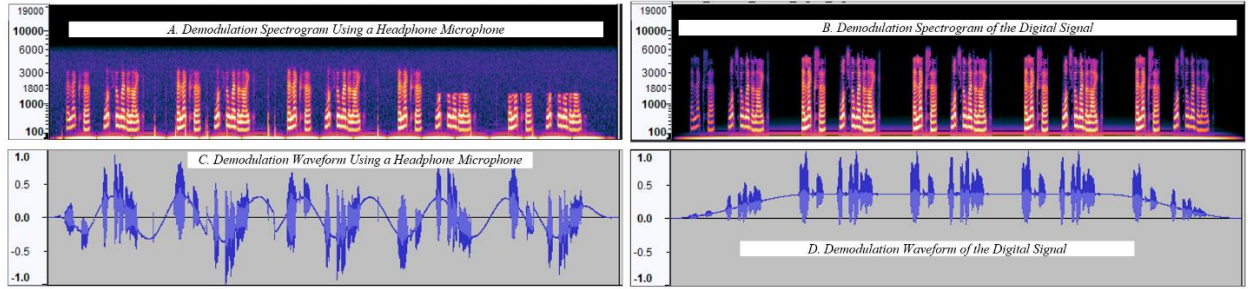

*Figure 8. Recovery of Inaudible Signal Aids Defensive Detection of Command Intention*

Running the demodulation algorithm on the digital variant of the attack signal reveals near-perfect recovery. The waveform of the demodulated signal captured by an inexpensive microphone indicates a periodic signal underneath various spikes in amplitude representing the original voice signal. Demodulating the digital signal reveals the upper portion of the envelope of the Tukey Window (Figure 8D).

## 4. DISCUSSION

The main results show that the attack method generates near-ultrasonic audio from audible sources using SUSBAM. IIn other words, the selective removal of modulated elements that might leak provides the spectral room to span the minimum detection needs of the non-linear microphone interactions. We demonstrate reversibility or demodulation of the inaudible signal to highlight future alerting methods and show that an attacker may embed secret (or subliminal) messages similar to audio steganography.

Several stages of SUSBAM might hinder its usefulness for microphone detection if compression, speed, pitch, or amplitude are incorrect. Proper compression, speed, pitch, and amplitude handling are crucial for generating near-ultrasonic audio that microphones can accurately detect and demodulate. Factors such as the low-pass filter, normalization, carrier frequency, SUSBAM modulation, and Tukey window can significantly affect the effectiveness of the near-ultrasonic audio transmission if not configured correctly. If the cutoff frequency is set too low, important frequency components might be filtered out, affecting the pitch and intelligibility of the signal. This lower cutoff might make it difficult for microphones to detect and recognize the original audio content. The normalization stage scales the signal to have a maximum amplitude of 1. If the original signal has a very low amplitude, the normalization could amplify noise and other undesired components, which might negatively affect microphone detection.

The choice of the carrier frequency (16 kHz in this example) is crucial for shifting the signal into the near-ultrasonic range. If the carrier frequency is too low, the modulated signal might still be within the audible range, causing interference with other sounds and reducing the effectiveness of the near-ultrasonic audio transmission. On the other hand, if the carrier frequency is too high, some microphones might not detect the modulated signal, as a low pass filter is generally employed to filter out frequencies greater than the human audible range (such as 20kHz).

If the input signal's amplitude or pitch is unsuitable for the chosen modulation scheme, the resulting modulated signal might be complex for microphones to detect or demodulate accurately. In addition, if the speed or pitch of the input signal is significantly altered before SUSBAM modulation, it might affect the modulated signal's ability to be accurately demodulated and recognized.

The Tukey window is applied to reduce spectral leakage and prevent artifacts in the audio signal. However, if the window's alpha parameter is not chosen correctly, it might cause undesirable effects on the signal, such as attenuating important frequency components or altering the pitch, which could hinder microphone detection.

Several advanced attack scenarios seem plausible, given that an attacker embeds the inaudible commands into silent periods of a well-attended public event. Previous work [19] has considered the potentially viral capabilities to embed NUIT signals in YouTube videos. One can consider future defensive actions by following the basic MITRE steps for the propagation of a denial-of-service or self-propagating worm attack. For example, an attacker could take stealth control of a public announcer system at a stadium or Grand Central Station to broadcast an inaudible trigger during a silent part, potentially reaching a large audience and their devices simultaneously. In both scenarios, the PA broadcasts an inaudible trigger to a large audience during a quiet part of the broadcast. To accomplish a NUIT attack, the attacker and defenders might address the following steps proactively, as described in Table 4.

*Table 4. NUIT-N Scenario Development for One-to-Many Attack and Defend Cycles*

| Stage | Attack Description | Plausible Defenses or Limits |
|---|---|---|
| **Prepare the inaudible trigger** | Utilize the SUSBAM technique or a similar method to create an inaudible audio command that triggers the desired response in the targeted devices. Ensure that the modulated signal is in the near-ultrasonic frequency range (e.g., 16-22 kHz) to avoid audible disturbances to the audience. | Advanced preparations may be needed, including intercepting a public announcer system control. Children under 14 may hear the annoying signal inaudible to those older in a mixed audience. |
| **PA system setup** | For the stadium scenario, install a high-quality public address system with speakers strategically placed throughout the venue to ensure even coverage. The design should be capable of accurately reproducing the inaudible trigger signal at a sufficient volume level. In the case of Grand Central Station, the PA system should be adapted to the train departure areas to ensure that the signal reaches all targeted devices. | The microphone-broadcast pairing appears as less than 60% reliable in initial trials. The event planning suggests a high risk for detection and a lower risk for success. |
| **Signal synchroni-zation** | To effectively broadcast the inaudible trigger, ensure proper synchronization of the PA system. Synchrony will help maintain signal integrity and prevent potential interference or audible artifacts during the silent part of the broadcast. | Previous batch modification of audio signals (using MATLAB or other advanced platforms) may not mix well with live streams. |
| **Timing** | Choose the appropriate timing for broadcasting the inaudible trigger, such as during a moment of silence at the stadium or a brief pause | If conceived as a prank or denial of service, the effects are, at best temporary. The introduction of remote control of |

|  | between train departure announcements at Grand Central Station. Timing control will increase the chances of the targeted devices receiving the signal without interference from other noises. | home devices such as locks or cameras may suggest consequential losses for those away from home but triggered by device attacks |
|---|---|---|
| **Testing** | Before broadcasting the inaudible trigger to the public, conduct tests in a controlled environment to ensure that the PA system can accurately reproduce the signal and that the targeted devices can detect and respond to it as intended | Validation and verification seem challenging before live events and inadequate trial attempts. Background noise and weather conditions may limit confidence even with tests. |

For securing devices, themselves, potential countermeasures at the user level could include acoustic shielding, frequency fingerprinting, user voice authentication, and device-level security updates, as summarized in Table 5.

*Table 5. Survey of Defensive Tactics to Detect Attacks and Protect Voice-Activated Devices*

| Defensive tactic | Description |
|---|---|
| **Acoustic shielding** | Designing devices with materials that dampen or reflect ultrasonic signals, reducing their impact on the device's microphone |
| **Frequency filtering** | Implementing digital filters to remove ultrasonic frequencies from the audio input, preventing the device from processing inaudible commands |
| **Machine learning** | Employing machine learning algorithms to identify and block unauthorized ultrasonic signals based on their unique characteristics |
| **User authentication** | Requiring user authentication before executing sensitive actions, thereby reducing the risk of unauthorized access or control |
| **Device-level security updates** | Ensuring that device manufacturers provide timely security updates and patches to address newly discovered vulnerabilities |

Future work will explore the device combinations and underlying non-linear physics associated with the device's particular broadcaster and receiver in tandem. Some limitations of the SUSBAM attack include the proximity requirement of the attacking device and the spatial requirement of the receiving device, which involves the geometry of the acoustical setting without deflecting ultrasonic wave surfaces. Additionally, specific speaker dependencies or biometrics, such as Siri voice print and Alexa profile for understanding household voices, may not entirely halt the attack but can act as a filter for specific dangerous commands. Finally, proper signal pre-processing is necessary to avoid audible leakage, as NUIT [19] suggests only babies can hear above 20 kHz. However, in some experiments, children under 14 were disturbed by the high-frequency attack, indicating the importance of addressing audible leakage even at 16 kHz thresholds.

To debug the near ultrasound commands that do not trigger executable actions, the Amazon Voice History [85] provides comparable recordings after the microphone and server have post-processed the inaudible to audible translations. While the present work examines multiple device [86] attack surface, future work [87] can systematically explore other hardware-software combinations.

## 5. CONCLUSIONS

In conclusion, the increasing prevalence of IoT devices in our daily lives presents significant security challenges. As this paper has shown, near ultrasonic attacks can seriously threaten these devices' security. As IoT technology continues to evolve, security researchers and practitioners must remain vigilant in identifying and mitigating emerging threats, such as near ultrasonic attacks, to ensure the safety and privacy of IoT users. Our findings replicate that ultrasonic attacks on voice-activated devices represent a viable threat with many potential consequences and extend previous work to include the MITRE security frameworks. We extend the attack surface to consider NUIT-N configurations where an attacking device simultaneously gains control over many (N) devices. The susceptibility of these devices to inaudible signals highlights the importance of developing countermeasures to protect users' authentication, privacy, and security.

## ACKNOWLEDGMENTS

The authors thank the PeopleTec Technical Fellows program for encouragement and project assistance.

## REFERENCES

[1] Wardino, J. (2023), Voice Search Statistics: Smart Speakers, Voice Assistants, and Users in 2023. https://serpwatch.io/blog/voice-search-statistics/

[2] Statista (2023), number of digital voice assistants in use worldwide from 2019 to 2024 (in billions), https://www.statista.com/statistics/973815/worldwide-digital-voice-assistant-in-use/

[3] Gritzalis, D., & Stergiopoulos, G. (2020). Unconventional Cyber Attacks: Facts-Not Fakes. In NATO Cyber Operations Seminar.

[4] Mentens, N. (2022). FOCUS: Frequency Based Detection of Covert Ultrasonic Signals. In ICT Systems Security and Privacy Protection: 37th IFIP TC11 International Conference, SEC 2022, Copenhagen, Denmark, June 13-15, 2022: Proceedings (Vol. 648, p. 70). Springer Nature.

[5] Iijima, R., Takehisa, T., & Mori, T. (2022, May). Cyber-physical firewall: monitoring and controlling the threats caused by malicious analog signals. In Proceedings of the 19th ACM International Conference on Computing Frontiers (pp. 296-304).

[6] Chen, Y., Gao, M., Liu, Y., Liu, J., Xu, X., Cheng, L., & Han, J. (2021). Implement of a secure selective ultrasonic microphone jammer. CCF Transactions on Pervasive Computing and Interaction, 3, 367-377.

[7] Silverajan, B., Ocak, M., & Nagel, B. (2018, July). Cybersecurity attacks and defences for unmanned smart ships. In 2018 IEEE international conference on internet of things (iThings) and IEEE green computing and communications (GreenCom) and IEEE cyber, physical and social computing (CPSCom) and IEEE smart data (SmartData) (pp. 15-20). IEEE.

[8] He, Y., Bian, J., Tong, X., Qian, Z., Zhu, W., Tian, X., & Wang, X. (2019, October). Canceling inaudible voice commands against voice control systems. In The 25th Annual International Conference on Mobile Computing and Networking (pp. 1-15).

[9] Huang, D., Tian, Z., Su, S., & Jiang, Y. (2020, December). A defense scheme of voice control system against DolphinAttack. In Proceedings of the 2020 International Conference on Cyberspace Innovation of Advanced Technologies (pp. 136-142).

[10] Lit, Y., Kim, S., & Sy, E. (2021, January). A survey on Amazon Alexa attack surfaces. In 2021 IEEE 18th Annual Consumer Communications & Networking Conference (CCNC) (pp. 1-7). IEEE.

[11] Park, Y., Choi, H., Cho, S., & Kim, Y. G. (2019). Security analysis of smart speaker: security attacks and mitigation. Computers, Materials & Continua, 61(3), 1075-1090.

[12] Kwon, H., Yoon, H., & Park, K. W. (2019, November). POSTER: Detecting audio adversarial example through audio modification. In Proceedings of the 2019 ACM SIGSAC conference on computer and communications security (pp. 2521-2523).

[13] Wu, Y., Xu, X., Walker, P. R., Liu, J., Saxena, N., Chen, Y., & Yu, J. (2021, May). HVAC: Evading Classifier-based Defenses in Hidden Voice Attacks. In Proceedings of the 2021 ACM Asia Conference on Computer and Communications Security (pp. 82-94).

[14] Meng, Y., Zhu, H., & Shen, X. (2022). Literature Review of Security in Smart Home Network. Security in Smart Home Networks, 21-35.

[15] Panoff, M., Dutta, R. G., Hu, Y., Yang, K., & Jin, Y. (2021). On sensor security in the era of IoT and CPS. SN Computer Science, 2(1), 51.

[16] Ponticello, A. (2020). Towards secure and usable authentication for voice-controlled smart home assistants (Doctoral dissertation, Wien).

[17] Stankovic, J. A., & Davidson, J. (2019). Raising awareness of security challenges for the internet of trillions of things. NAE Bridge Magazine, 49(3), 40-45.

[18] Gong, Y., & Poellabauer, C. (2018). An overview of vulnerabilities of voice controlled systems. arXiv preprint arXiv:1803.09156.

[19] Xia, Q., Chen, Q., Xu, S. (2023) Near-Ultrasound Inaudible Trojan (NUIT): Exploiting Your Speaker to Attack Your Microphone, UseNix Security 2023, https://www.usenix.org/system/files/sec23fall-prepub-261-xia-qi.pdf

[20] Schönherr, L., Golla, M., Eisenhofer, T., Wiele, J., Kolossa, D., & Holz, T. (2020). Unacceptable, where is my privacy? exploring accidental triggers of smart speakers. arXiv preprint arXiv:2008.00508.

[21] Oh, T., Aiken, W., & Kim, H. (2018, July). Hey Siri–Are You There?: Jamming of Voice Commands Using the Resonance Effect (Work-in-Progress). In 2018 International Conference on Software Security and Assurance (ICSSA) (pp. 73-76). IEEE.

[22] Chen, Y. (2020). Practical Adversarial Attacks Against Black Box Speech Recognition Systems and Devices (Doctoral dissertation, Florida Institute of Technology).

[23] Cho, G., Choi, J., Kim, H., Hyun, S., & Ryoo, J. (2019). Threat modeling and analysis of voice assistant applications. In Information Security Applications: 19th International Conference, WISA 2018, Jeju Island, Korea, August 23–25, 2018, Revised Selected Papers 19 (pp. 197-209). Springer International Publishing.

[24] Meng, Y., Zhang, W., Zhu, H., & Shen, X. S. (2018). Securing consumer IoT in the smart home: Architecture, challenges, and countermeasures. IEEE Wireless Communications, 25(6), 53-59.

[25] Pathak, S., Islam, S. A., Jiang, H., Xu, L., & Tomai, E. (2022). A survey on security analysis of Amazon echo devices. High-Confidence Computing, 100087.

[26] Li, L., Liu, M., Yao, Y., Dang, F., Cao, Z., & Liu, Y. (2020, November). Patronus: Preventing unauthorized speech recordings with support for selective unscrambling. In Proceedings of the 18th Conference on Embedded Networked Sensor Systems (pp. 245-257).

[27] Zhang, S., & Das, A. (2021, October). HandLock: Enabling 2-FA for Smart Home Voice Assistants using Inaudible Acoustic Signal. In Proceedings of the 24th International Symposium on Research in Attacks, Intrusions and Defenses (pp. 251-265).

[28] Zhang, N., Mi, X., Feng, X., Wang, X., Tian, Y., & Qian, F. (2019, May). Dangerous skills: Understanding and mitigating security risks of voice-controlled third-party functions on virtual personal assistant systems. In 2019 IEEE Symposium on Security and Privacy (SP) (pp. 1381-1396). IEEE.

[29] Kasher, M., Zhao, M., Greenberg, A., Gulati, D., Kokalj-Filipovic, S., & Spasojevic, P. (2021, June). Inaudible Manipulation of Voice-Enabled Devices Through BackDoor Using Robust Adversarial Audio Attacks. In Proceedings of the 3rd ACM Workshop on Wireless Security and Machine Learning (pp. 37-42).

[30] Koffas, S., Xu, J., Conti, M., & Picek, S. (2021). Can you hear it? backdoor attacks via ultrasonic triggers. arXiv preprint arXiv:2107.14569.

[31]   Du, T., Ji, S., Li, J., Gu, Q., Wang, T., & Beyah, R. (2020, October). Sirenattack: Generating adversarial audio for end-to-end acoustic systems. In Proceedings of the 15th ACM Asia Conference on Computer and Communications Security (pp. 357-369).

[32]   Song, L., & Mittal, P. (2017, October). Poster: Inaudible voice commands. In Proceedings of the 2017 ACM SIGSAC Conference on Computer and Communications Security (pp. 2583-2585).

[33]   Wang, Y., Guo, H., & Yan, Q. (2022). Ghosttalk: Interactive attack on smartphone voice system through power line. arXiv preprint arXiv:2202.02585.

[34]   Park, S. H., & Lee, I. G. (2020, November). Effective voice fuzzing method for finding vulnerabilities in ai speech recognition devices. In 2020 IEEE International Conference on Intelligence and Security Informatics (ISI) (pp. 1-6). IEEE.

[35]   Alzantot, M., Balaji, B., & Srivastava, M. (2018). Did you hear that? adversarial examples against automatic speech recognition. arXiv preprint arXiv:1801.00554.

[36]   Maji, R., Biswas, A., & Chaki, R. (2022, November). A Look into the Vulnerability of Voice Assisted IoT. In Computer Information Systems and Industrial Management: 21st International Conference, CISIM 2022, Barranquilla, Colombia, July 15–17, 2022, Proceedings (pp. 49-62). Cham: Springer International Publishing.

[37]   Bolton, T., Dargahi, T., Belguith, S., Al-Rakhami, M. S., & Sodhro, A. H. (2021). On the security and privacy challenges of virtual assistants. Sensors, 21(7), 2312.

[38]   Zhang, G., Yan, C., Ji, X., Zhang, T., Zhang, T., & Xu, W. (2017, October). Dolphinattack: Inaudible voice commands. In Proceedings of the 2017 ACM SIGSAC conference on computer and communications security (pp. 103-117).

[39]   Yan, C., Zhang, G., Ji, X., Zhang, T., Zhang, T., & Xu, W. (2019). The feasibility of injecting inaudible voice commands to voice assistants. IEEE Transactions on Dependable and Secure Computing, 18(3), 1108-1124.

[40]   Roy, N., Shen, S., Hassanieh, H., & Choudhury, R. R. (2018). Inaudible voice commands: The long-range attack and defense. In 15th {USENIX} Symposium on Networked Systems Design and Implementation ({NSDI} 18) (pp. 547-560).

[41]   Zhang, G., Ji, X., Li, X., Qu, G., & Xu, W. (2021, February). EarArray: Defending against DolphinAttack via Acoustic Attenuation. In NDSS.

[42]   Zhou, M., Qin, Z., Lin, X., Hu, S., Wang, Q., & Ren, K. (2019). Hidden voice commands: Attacks and defenses on the VCS of autonomous driving cars. IEEE Wireless Communications, 26(5), 128-133.

[43]   Doi, K., & Sugawara, T. (2022, November). Poster: Inaudible Acoustic Noise from Silicon Capacitors for Voice-Command Injection. In Proceedings of the 2022 ACM SIGSAC Conference on Computer and Communications Security (pp. 3339-3341).

[44]   Yuan, X., Chen, Y., Wang, A., Chen, K., Zhang, S., Huang, H., & Molloy, I. M. (2018, December). All your Alexa are belong to us: A remote voice control attack against echo. In 2018 IEEE global communications conference (GLOBECOM) (pp. 1-6). IEEE.

[45]   Mao, J., Zhu, S., & Liu, J. (2020). An inaudible voice attack to context-based device authentication in smart IoT systems. Journal of Systems Architecture, 104, 101696.

[46]   Gao, C., Chandrasekaran, V., Fawaz, K., & Banerjee, S. (2018, August). Traversing the quagmire that is privacy in your smart home. In Proceedings of the 2018 Workshop on IoT Security and Privacy (pp. 22-28).

[47]   Zhang, G. Yan C., Ji X., Zhang T., Zhang T., Xu W. DolphinAttack: Inaudible Voice Commands. In Proceedings of the 2017 ACM SIGSAC Conference on Computer and Communications Security (Dallas, TX, USA, 2017).

[48]     Arumugam, S. D., Hari Prashanth, S., Laksminarayanan, G., & Ananthi, N. (2020). Voice assistants through inaudible voice commands for visually challenged people using gesture algorithm. European Journal of Molecular & Clinical Medicine, 7(4), 2020.

[49]     Lien, J., Al Momin, M. A., & Yuan, X. (2022). Attacks on Voice Assistant Systems. In Security, Data Analytics, and Energy-Aware Solutions in the IoT (pp. 61-77). IGI Global.

[50]     Gong, Y., & Poellabauer, C. (2018, July). Protecting voice controlled systems using sound source identification based on acoustic cues. In 2018 27th International Conference on Computer Communication and Networks (ICCCN) (pp. 1-9). IEEE.

[51]     Iijima, R., Minami, S., Yunao, Z., Takehisa, T., Takahashi, T., Oikawa, Y., & Mori, T. (2018, October). Audio hotspot attack: An attack on voice assistance systems using directional sound beams. In Proceedings of the 2018 ACM SIGSAC Conference on Computer and Communications Security (pp. 2222-2224).

[52]     Chen, T., Shangguan, L., Li, Z., & Jamieson, K. (2020, January). Metamorph: Injecting inaudible commands into over-the-air voice controlled systems. In Network and Distributed Systems Security (NDSS) Symposium.

[53]     Chen, Y., Yuan, X., Zhang, J., Zhao, Y., Zhang, S., Chen, K., & Wang, X. (2020, August). Devil's Whisper: A General Approach for Physical Adversarial Attacks against Commercial Black-box Speech Recognition Devices. In USENIX Security Symposium (pp. 2667-2684).

[54]     McCarthy, A., Gaster, B. R., & Legg, P. (2020, June). Shouting Through Letterboxes: A study on attack susceptibility of voice assistants. In 2020 International Conference on Cyber Security and Protection of Digital Services (Cyber Security) (pp. 1-8). IEEE.

[55]     Li, Z., Shi, C., Zhang, T., Xie, Y., Liu, J., Yuan, B., & Chen, Y. (2021, November). Robust detection of machine-induced audio attacks in intelligent audio systems with microphone array. In Proceedings of the 2021 ACM SIGSAC Conference on Computer and Communications Security (pp. 1884-1899).

[56]     Shahrad, M., Mosenia, A., Song, L., Chiang, M., Wentzlaff, D., & Mittal, P. (2017). Acoustic denial of service attacks on HDDs. arXiv preprint arXiv:1712.07816.

[57]     Lenhardt, M. L., Skellett, R., Wang, P., & Clarke, A. M. (1991). Human ultrasonic speech perception. Science, 253(5015), 82-85.

[58]     Akiyama, M. (2010). Silent alarm: The mosquito youth deterrent and the politics of frequency. Canadian Journal of Communication, 35(3), 455-471.

[59]     Schönherr, L., Kohls, K., Zeiler, S., Holz, T., & Kolossa, D. (2018). Adversarial attacks against automatic speech recognition systems via psychoacoustic hiding. *arXiv preprint arXiv:1808.05665*.

[60]     Mao, J., Zhu, S., Dai, X., Lin, Q., & Liu, J. (2020). Watchdog: Detecting ultrasonic-based inaudible voice attacks to smart home systems. IEEE Internet of Things Journal, 7(9), 8025-8035.

[61]     El-Rewini, Z., Sadatsharan, K., Sugunaraj, N., Selvaraj, D. F., Plathottam, S. J., & Ranganathan, P. (2020). Cybersecurity attacks in vehicular sensors. IEEE Sensors Journal, 20(22), 13752-13767.

[62]     Xu, W., Yan, C., Jia, W., Ji, X., & Liu, J. (2018). Analyzing and enhancing the security of ultrasonic sensors for autonomous vehicles. IEEE Internet of Things Journal, 5(6), 5015-5029.

[63]     Silverajan, B., Ocak, M., & Nagel, B. (2018, July). Cybersecurity attacks and defences for unmanned smart ships. In 2018 IEEE international conference on internet of things (iThings) and IEEE green computing and communications (GreenCom) and IEEE cyber, physical and social computing (CPSCom), and IEEE smart data (SmartData) (pp. 15-20). IEEE.

[64]     Gluck, T., Kravchik, M., Chocron, S., Elovici, Y., & Shabtai, A. (2020). Spoofing attack on ultrasonic distance sensors using a continuous signal. Sensors, 20(21), 6157.

[65]     Guri, M., Solewicz, Y., & Elovici, Y. (2018, December). Mosquito: Covert ultrasonic transmissions between two air-gapped computers using speaker-to-speaker communication. In 2018 IEEE Conference on Dependable and Secure Computing (DSC) (pp. 1-8). IEEE.

[66] Yan, Q., Liu, K., Zhou, Q., Guo, H., & Zhang, N. (2020, February). Surfingattack: Interactive hidden attack on voice assistants using ultrasonic guided waves. In Network and Distributed Systems Security (NDSS) Symposium.

[67] Guri, M., Solewicz, Y., & Elovici, Y. (2020). Speaker-to-speaker covert ultrasonic communication. Journal of Information Security and Applications, 51, 102458.

[68] Cheek, E., Khuttan, D., Changalvala, R., & Malik, H. (2020, December). Physical fingerprinting of ultrasonic sensors and applications to sensor security. In 2020 IEEE 6th International Conference on Dependability in Sensor, Cloud and Big Data Systems and Application (DependSys) (pp. 65-72). IEEE.

[69] Sherry, R., Bayne, E., & McLuskie, D. (2023, March). Near-Ultrasonic Covert Channels Using Software-Defined Radio Techniques. In Proceedings of the International Conference on Cybersecurity, Situational Awareness and Social Media: Cyber Science 2022; 20–21 June; Wales (pp. 169-189). Singapore: Springer Nature Singapore.

[70] Gao, M., Chen, Y., Li, Y., Zhang, L., Liu, J., Lu, L., ... & Ren, K. (2023). A Resilience Evaluation Framework on Ultrasonic Microphone Jammers. IEEE Transactions on Mobile Computing.

[71] McGhee, J. E. (2016). Liberating cyber offense. Strategic Studies Quarterly, 10(4), 46-63.

[72] Edwards, A. O. (2016). Ultrasonic Data Steganography.

[73] Wong, W. (2018). *Crossing the Air Gap—An Ultrasonic Covert Channel* (Doctoral dissertation).

[74] Mentens, N. (2022). FOCUS: Frequency-Based Detection of Covert Ultrasonic Signals. In ICT Systems Security and Privacy Protection: 37th IFIP TC11 International Conference, SEC 2022, Copenhagen, Denmark, June 13-15, 2022: Proceedings (Vol. 648, p. 70). Springer Nature

[75] Guri, M. (2021, December). GAIROSCOPE: Leaking Data from Air-Gapped Computers to Nearby Smartphones using Speakers-to-Gyro Communication. In 2021 18th International Conference on Privacy, Security and Trust (PST) (pp. 1-10). IEEE.

[76] Chen, Y., Gao, M., Liu, Y., Liu, J., Xu, X., Cheng, L., & Han, J. (2021). Implement a secure selective ultrasonic microphone jammer. CCF Transactions on Pervasive Computing and Interaction, 3, 367-377.

[77] Iijima, R., Takehisa, T., & Mori, T. (2022, May). Cyber-physical firewall: monitoring and controlling the threats caused by malicious analog signals. In Proceedings of the 19th ACM International Conference on Computing Frontiers (pp. 296-304).

[78] Kwon, R., Ashley, T., Castleberry, J., Mckenzie, P., & Gourisetti, S. N. G. (2020, October). Cyber threat dictionary using MITRE ATT&CK matrix and NIST cybersecurity framework mapping. In 2020 Resilience Week (RWS) (pp. 106-112). IEEE.

[79] Strom, B. E., Applebaum, A., Miller, D. P., Nickels, K. C., Pennington, A. G., & Thomas, C. B. (2018). Mitre att&ck: Design and philosophy. In *Technical report*. The MITRE Corporation.

[80] Al-Shaer, R., Spring, J. M., & Christou, E. (2020, June). Learning the associations of MITRE ATT&CK adversarial techniques. In 2020 IEEE Conference on Communications and Network Security (CNS) (pp. 1-9). IEEE.

[81] Kaloroumakis, P. E., & Smith, M. J. (2021). Toward a knowledge graph of cybersecurity countermeasures. *The MITRE Corporation*, 11.

[82] Kim, K., Shin, Y., Lee, J., & Lee, K. (2021). Automatically attributing mobile threat actors by vectorized ATT&CK matrix and paired indicator. Sensors, 21(19), 6522.

[83] Lakshmanan, R. (2021), Malicious Amazon Alexa Skills Can Easily Bypass Vetting Process, Hacker News, https://thehackernews.com/2021/02/alert-malicious-amazon-alexa-skills-can.html

[84] McKee, F., Noever,D. (2023). Online video demonstrations. Weather Example, Android-Echo Dot Gen 2, https://deeperbrain.com/us/weather_trojan.m4v; Wikipedia information search, Android-Echo Dot Gen 3,

[https://deeperbrain.com/us/wikipedia_trojan_dot3.mp4](https://deeperbrain.com/us/wikipedia_trojan_dot3.mp4), NUIT-N Weather Example, Android to Both Echo Dots Gen 2/3, [https://deeperbrain.com/us/nuitn.mp4](https://deeperbrain.com/us/nuitn.mp4)

[85] Amazon Review Voice History (2023), [https://www.amazon.com/alexa-privacy/apd/rvh](https://www.amazon.com/alexa-privacy/apd/rvh)

[86] Alexa Device Software Versions (2023), The vulnerabilities were specifically tested here for two software and hardware versions, although [19] has demonstrated a wider array of devices that make the vulnerability common to all the major microphones. Amazon Echo (2nd Generation)-Latest Software Version: 8960323972; Amazon Echo (3rd Generation) Latest Software Version: 8960323972 [https://www.amazon.com/gp/help/customer/display.html?nodeId=GMB5FVUB6REAVTXY](https://www.amazon.com/gp/help/customer/display.html?nodeId=GMB5FVUB6REAVTXY)

[87] Github (2023), Hardware and software matrix updated for near ultrasonic attack testing [https://github.com/reveondivad/nuance/blob/main/NearUltraTests.csv](https://github.com/reveondivad/nuance/blob/main/NearUltraTests.csv)

## Authors

**Forrest McKee** has AI research experience with the Department of Defense in object detection and reinforcement learning. He received his Bachelor's (BS) and Master's (MSE) from the University of Alabama, Huntsville, Engineering.

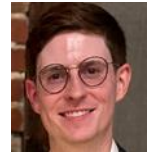

**David Noever** has research experience with NASA and the Department of Defense in machine learning and data mining. He received his BS from Princeton University and his Ph.D. from Oxford University, as a Rhodes Scholar, in theoretical physics.

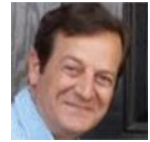

### Appendix A: Example Stages of MITRE Tactic and Technique Classifications in both Attack and Defend Cycles with Reference to Viable Near Ultrasonic Methods

| ATT&CK Tactic | ATT&CK Technique | D3FEND Tactic | D3FEND Technique | Ultrasonic Attack |
|---|---|---|---|---|
| **Initial Access** | T1189: Drive-by Compromise | User Training | D3-T1023: Security Awareness Training | Yes |
|  | T1200: Hardware Additions | Physical Security | D3-T1042: Physical Access Control | Yes |
| **Execution** | T1204: User Execution | User Training | D3-T1023: Security Awareness Training | Yes |
|  | T1129: Execution through Module Load | Application Control | D3-T1008: Application Whitelisting | Yes |
| **Persistence** | T1060: Registry Run Keys / Startup Folder | System Hardening | D3-T1035: Security Policy Enforcement | Yes |
|  | T1166: Setuid and Setgid | System Hardening | D3-T1005: Privilege Management | Yes |
| **Privilege Escalation** | T1078: Valid Accounts | Access Control | D3-T1021: User Account Management | Yes |
|  | T1134: Access Token Manipulation | System Hardening | D3-T1005: Privilege Management | Yes |
| **Defense Evasion** | T1140: Deobfuscate/ Decode Files or Information | Malware Protection | D3-T1009: Anti-Malware Deployment | Yes |

| ATT&CK Tactic | ATT&CK Technique | D3FEND Tactic | D3FEND Technique | Ultrasonic Attack |
|---|---|---|---|---|
| | T1158: Hidden Files and Directories | Logging & Monitoring | D3-T1016: Log Management | Yes |
| **Credential Access** | T1056: Input Capture | User Training | D3-T1023: Security Awareness Training | Yes |
| | T1110: Brute Force | Account Lockout Policy | D3-T1022: Account Lockout Policy Implementation | Yes |
| **Lateral Movement** | T1021: Remote Services | Access Control | D3-T1021: Service Account Management | Yes |
| | T1028: Windows Remote Management | Network Segmentation | D3-T1041: Network Segmentation | Yes |
| **Collection** | T1113: Screen Capture | User Training | D3-T1023: Security Awareness Training | Yes |
| | T1119: Automated Collection | Data Loss Prevention | D3-T1010: Data Classification | Yes |
| **Command and Control** | T1043: Commonly Used Port | Network Segmentation | D3-T1041: Network Segmentation | Yes |
| | T1573: Encrypted Channel | Network Monitoring | D3-T1017: Network Traffic Analysis | Yes |
| **Exfiltration** | T1041: Exfiltration Over Command and Control Channel | Data Loss Prevention | D3-T1010: Data Classification | Yes |
| | T1567: Exfiltration Over Web Service | Network Monitoring | D3-T1017: Network Traffic Analysis | Yes |